\documentclass[12pt,preprint]{aastex}

\shorttitle{Variability of GRB 021004}
\shortauthors{Bersier et al.}

\begin{document}

\title{The unusual optical afterglow of the gamma-ray burst GRB
021004: Color changes and short-time-scale
variability\altaffilmark{1}}

\author{D.~Bersier\altaffilmark{2}, 
K.~Z.~Stanek\altaffilmark{2},
J.~Winn\altaffilmark{2,3},
T.~Grav\altaffilmark{2,4},
M.~J.~Holman\altaffilmark{2},
T.~Matheson\altaffilmark{2},
B.~Mochejska\altaffilmark{2,5}
D. Steeghs\altaffilmark{2},
A.~R.~Walker\altaffilmark{6},
P.~M.~Garnavich\altaffilmark{7},
J.~Quinn\altaffilmark{7},
S.~Jha\altaffilmark{8},
H.~Calitz\altaffilmark{9},
P.~Meintjes\altaffilmark{9}
}

\altaffiltext{1}{Based on data from the FLWO 1.2m telescope, the
Magellan 6.5m Landon Clay telescope, the CTIO Blanco 4m telescope, the
1.8m Vatican Advanced Technology Telescope, and the Boyden 1.52m
telescope.}

\altaffiltext{2}{Harvard-Smithsonian Center for Astrophysics, 60
Garden St., Cambridge, MA 02138}

\altaffiltext{3}{NSF Astronomy \& Astrophysics Postdoctoral Fellow}

\altaffiltext{4}{Institute of Theoretical Astrophysics, University in
Oslo, Norway}

\altaffiltext{5}{Hubble Fellow}

\altaffiltext{6}{Cerro Tololo Inter-American Observatory, National
Optical Astronomy Observatory, Casilla 603, La Serena, Chile.  NOAO is
operated by AURA Inc., under cooperative agreement with the National
Science Foundation.}

\altaffiltext{7}{Department of Physics, University of Notre Dame,
Notre Dame, IN 46556-5670}

\altaffiltext{8}{Astronomy Department, University of California at
Berkeley, Berkeley, CA 94720-3411}

\altaffiltext{9}{University of the Free State, South Africa}

\email{dbersier@cfa.harvard.edu, 
kstanek@cfa.harvard.edu,
jwinn@cfa.harvard.edu,
tgrav@cfa.harvard.edu,
mholman@cfa.harvard.edu,
tmatheson@cfa.harvard.edu,
bmochejska@cfa.harvard.edu,
dsteeghs@head-cfa.harvard.edu,
awalker@noao.edu,
pgarnavi@miranda.phys.nd.edu,
jquinn@miranda.phys.nd.edu,
saurabh@astron.berkeley.edu,
MeintjPJ@sci.uovs.ac.za }

\begin{abstract}

We report $UBVRI$ observations of the optical afterglow of the
gamma-ray burst GRB 021004. We observed significant ($\sim 10-20$\%)
deviations from a power law decay on several time scales, ranging from
a few hours down to 20-30 minutes. We also observed a significant
color change starting $\sim 1.5$ days after the burst, confirming and
extending the spectroscopic results already reported by Matheson et
al. (2002).  We discuss these results in the context of several models
that have recently been proposed to account for the anomalous
photometric behavior of this event.

\end{abstract}


\section{Introduction}
\label{sect_intro}

The gamma ray burst GRB 021004 was discovered by HETE at 12:06 UT on
4~October 2002 \citep{gcn1565}. Observations beginning less than 10
minutes after the burst revealed a bright fading source
\citep{gcn1564} located at $\alpha_{2000} = 00^h 26^m 54\fs 7,\
\delta_{2000} = +18\arcdeg 55\arcmin 42\arcsec$ \citep{gcn1592}.  A
radio counterpart was found at 22.5 GHz \citep{gcn1574}, 15 GHz
\citep{gcn1588} and at 86 GHz \citep{gcn1590}. Polarimetric
observations were performed (Covino et al. 2002, Rol et al. 2002).  A
spectrum showed that the redshift was $z\geq 2.3$ \citep{gcn1605},
later refined to 2.3351 \citep{moller02}.  In addition an X-ray
afterglow was found \citep{gcn1624}.


Several things set this burst apart from other bursts for which
afterglows have been observed.  Radio observations revealed that the
radio afterglow had a very unusual spectrum \citep{gcn1612}.  Optical
spectra showed several absorption line systems, some being separated
from the presumed host galaxy by $\sim 3000\ \mbox{km} \,
\mbox{s}^{-1}$ (Chornock \& Filippenko 2002, Mirabal et al. 2002,
Matheson et al. 2002, M\o ller et al. 2002).  Furthermore, optical
spectra showed a significant change in the blue portion of the
spectrum whereas the red end did not change \citep{tom02}.  In
addition to this, the photometric behavior of the optical transient
(OT) was highly unusual. The optical afterglow faded quickly, and
seemed to exhibit a break \citep{gcn1573} but intensive monitoring
revealed that the fading was not as fast as expected (Winn et
al. 2002; Halpern et al. 2002a). The afterglow resumed fading
\citep{gcn1586} but stalled again after $\sim 2$ days \citep{gcn1598}.
There are also clear deviations from an expected power-law decay,
representing variability on short time scales.

Here we report our intensive photometry of the optical transient (OT)
during the first night, followed by more occasional monitoring over
the next few weeks.

\section{Observations}

\label{sect_phot}

Most of our $UBVR_C I_C$ data were obtained with the F. L. Whipple
Observatory (FLWO) 1.2m telescope equipped with the ``4Shooter''
mosaic camera which delivers a pixel scale of $0\farcs 335$ per
pixel. We continuously monitored the afterglow during the first night,
with a typical exposure time of $300$ sec. We also obtained several
measurements per night for the next five nights.  As the burst became
fainter, we continued obtaining data (Oct 9 -- Oct 12) with the
Magellan 6.5m Landon Clay telescope at Las Campanas Observatory, using
the Magellan Instant Camera (with a pixel scale is $0\farcs 069$ per
pixel). The typical exposure time for the Magellan observations was 10
minutes in each band.

We also obtained $BVRI$ data during the first night from the CTIO 4m
telescope with the MOSAIC II camera.  Another early $R$-band data
point comes from the Boyden 1.52m telescope (University of the Free
State, South Africa).  In addition we observed the optical transient
about 19 days after the burst with the Vatican Advanced Technology
Telescope (VATT) 1.8-m telescope.

The data were reduced by several of us using three different
photometry packages.  We used DoPHOT \citep{sms93}, DAOPHOT II
(Stetson, 1987, 1992; Stetson \& Harris 1988) and the image
subtraction code ISIS (Alard \& Lupton 1998, Alard 2000) and we found
excellent agreement between the various packages.  For consistency, we
present the photometry obtained with DAOPHOT throughout this paper.
Images were brought onto a common zero point using from 50 to 100
stars per image.  We used several stars described by \citet{gcn1630}
to calibrate the instrumental magnitudes, choosing blue stars with
colors similar to the OT. In particular, the star located at
$\alpha_{2000} = 00^h 26^m 51\fs 4,\ \delta_{2000} = +18\arcdeg
54\arcmin 36\arcsec$ is close to the OT and has very similar colors;
it is thus particularly useful for calibrating the photometry.

\section{Temporal behavior}
\label{sec_time}

\begin{figure}
\plotone{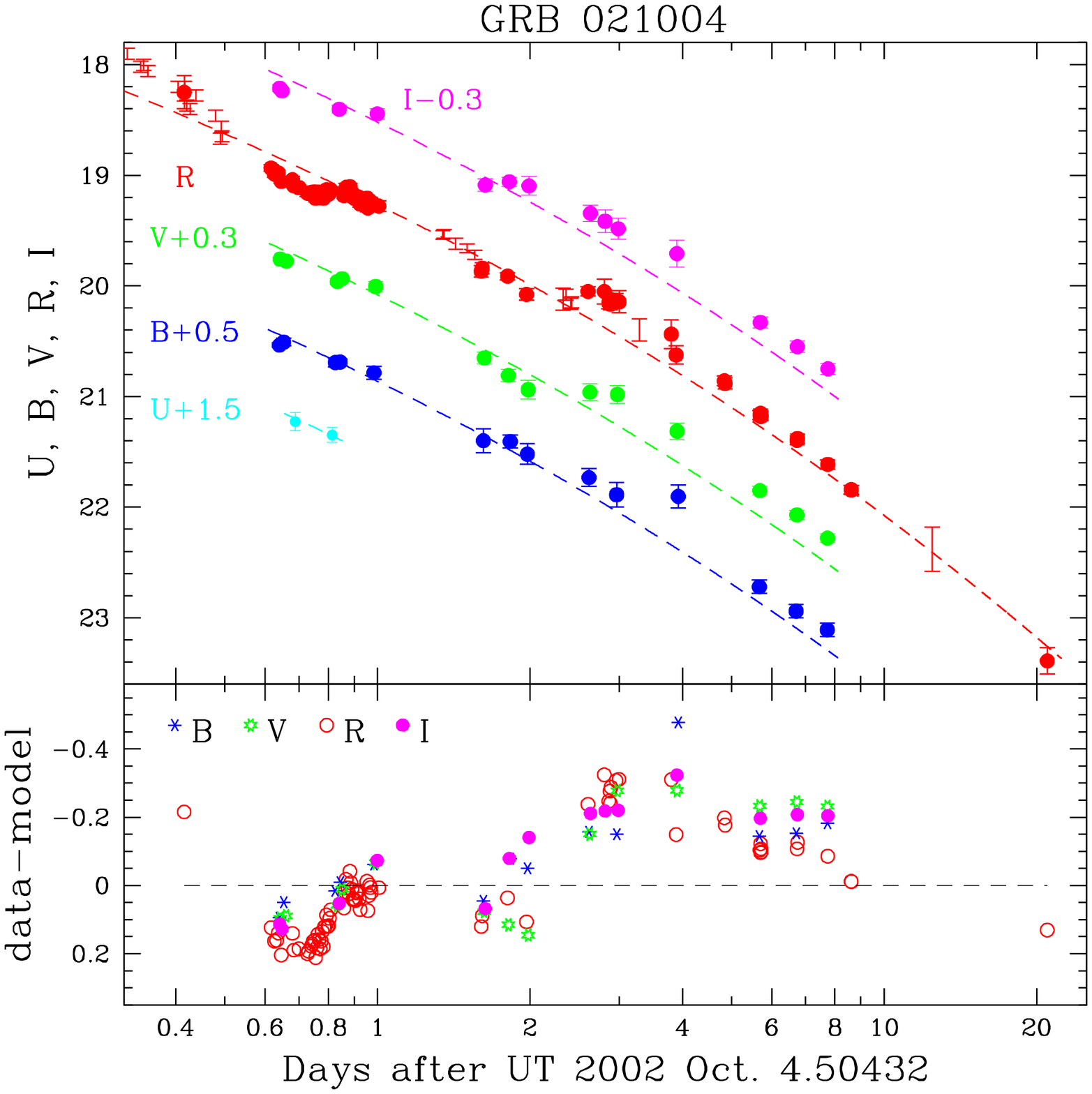}
\caption{{\it Upper panel:} $UBVRI$ light curves of GRB 021004. Our
data are shown with filled circles, data from the GCN are shown with
error bars only. We used $R$-band data from \citet{gcn1564},
\citet{gcn1566}, \citet{gcn1573}, \citet{gcn1585}, \citet{gcn1597},
\citet{gcn1587}, \citet{gcn1594}, \citet{gcn1603}, \citet{gcn1607},
\citet{gcn1645}, \citet{gcn1615}.  Also shown are the simple
analytical fits discussed in the text. {\it Lower panel:} residuals
(data$-$model) for our $BVRI$ data. One clearly sees the bumpy
character of the light curve as well as the color evolution of the
afterglow.}
\label{fig_fit}
\end{figure}

It was obvious from early on that the optical afterglow exhibited an
unusual behavior.  In the first hour after the burst, the OT faded
monotonically but this behavior almost stopped after 45 minutes. The
OT reached a secondary peak $\sim 2.5$ hours after the burst.  Fading
then resumed and the OT more closely followed the usual power law
decay now commonly observed in GRBs.

We plot the GRB 021004 $UBVRI$ light curves in the upper panel of
Fig.\ref{fig_fit}, omitting the first $\sim 10$ hours for clarity.
Most of the early $UBVRI$ data come from the FLWO 1.2-m telescope
(Winn et al. 2002; Bersier et al. 2002; Stanek et al. 2002), with one
early $R$-band point from the Boyden 1.52-m telescope and one $BVRI$
set from the CTIO 4-m telescope, and additional later data ($t>5.5$
days) from the 6.5-m Magellan and from the 1.8-m VATT telescopes
\citep{gcn1661}.  To obtain as clear a picture as possible of the
temporal evolution of the afterglow, we also display $R$-band data
taken earlier and in between our data, selecting when possible
uniformly reduced data sets as posted on the Gamma-Ray Burst
Coordinates Network (GCN; see caption of Fig.~\ref{fig_fit} for the
list of data sets we used).  To allow for small differences in the
reduction procedures and photometric calibration, uncertainties
smaller than $0.05$ mag in the GCN data were increased to $0.05$
mag. The combined data set has the following number of points:
$N(U,B,V,R,I) = (2,14,14,121,14)$, for a total of 165 points, of which
125 points are our own observations.

To describe the temporal evolution of the GRB 021004 optical
counterpart, we fitted the compiled $UBVRI$ data with the smoothly
broken power-law model of \citet{beuermann}:
\begin{equation}
F_{\nu}(t) =
\frac{2F_{\nu,0}}{\left[\left(\frac{t}{t_b}\right)^{\alpha_1 s}
+\left(\frac{t}{t_b}\right)^{\alpha_2 s}\right]^{1/s}} \ ,
\end{equation}
where $t_b$ is the time of the break, $F_{\nu,0}$ is the flux at $t_b$
and $s$ controls the sharpness of the break. This formula describes a
power-law $t^{-\alpha_1}$ decline at early times ($t\ll t_b$) and
another power-law $t^{-\alpha_2}$ decline at late times (for details
see Stanek et al. 2001).

The results of the combined fit are shown as the dashed lines in the
upper panel of Fig.\ref{fig_fit}. Clearly, the smooth model is a poor
fit to the data but provides a reasonable approximation of the general
trend. There are clear ``bumps and wiggles'' in all bands (except $U$
where we have only two measurements). Due to the poor fit, the
parameters obtained would be very different if only a subsample of the
data were fitted, or if the data were sampled differently.  With these
caveats, we report the best-fit values: $\alpha_1 = 0.5, \alpha_2 =
2.4, s\approx 0.3, t_b=14\;$days (we do not give errors on these
values). With those parameters the model fits the early ($\sim 0.01$
day) and late $R$-band data reasonably well.

In the lower panel of Fig.\ref{fig_fit} we show residuals
(data$-$model) for our $BVRI$ data. Here the bumpy character of the
light curve is obvious. What can be also seen is that the broad-band
colors of the OT were changing: while $VR$-bands remained constant to
$\sim 0.1\;$mag between the end of night 2 and during night 3 and then
decayed by $\sim 0.3$ mag when observed on night 4, $BI$-bands decayed
by $\sim 0.4$ mag between the end of night 2 and during night 3, with
no further decay in $B$-band when observed on night 4. In further
support of the claim for chromatic decay, it should be mentioned that
Rhoads, Burud \& Fruchter (2002) reported fading of $0.47\pm 0.04$ mag
in the $H$-band between nights 2 and 3, similar to the decay in the
$I$-band over the same time and very different from the behavior in
the $R$-band.

\section{Time evolution of the energy distribution}
\label{sec_spec}

This is the first observed clear example of an OT changing color as it
fades.  The observed change agrees with the one observed
spectroscopically for the same GRB afterglow by \citet{tom02}, in the
sense that between nights 1 and 3 the afterglow became redder (both
$B-V$ and $B-R$ increase). There were spectroscopic observations made
on nights beyond night 3 (e.g.~Chornock \& Filippenko 2002). We
predict that they will reveal a reverse change ($B-R$ decreasing when
comparing night 4 to night 3) since we see the energy distribution
come back to what it was on night 1.  We consider the detection of a
significant color changes in the OT of the GRB\,021004 to be very
secure\footnote{To allow the astronomical community to verify our
measurements independently, we have placed all of our data, including
individual CCD frames, on {\tt anonymous ftp} at {\tt
ftp://cfa-ftp.harvard.edu/pub/kstanek/GRB021004}.}.

\begin{figure}[h]
\plotone{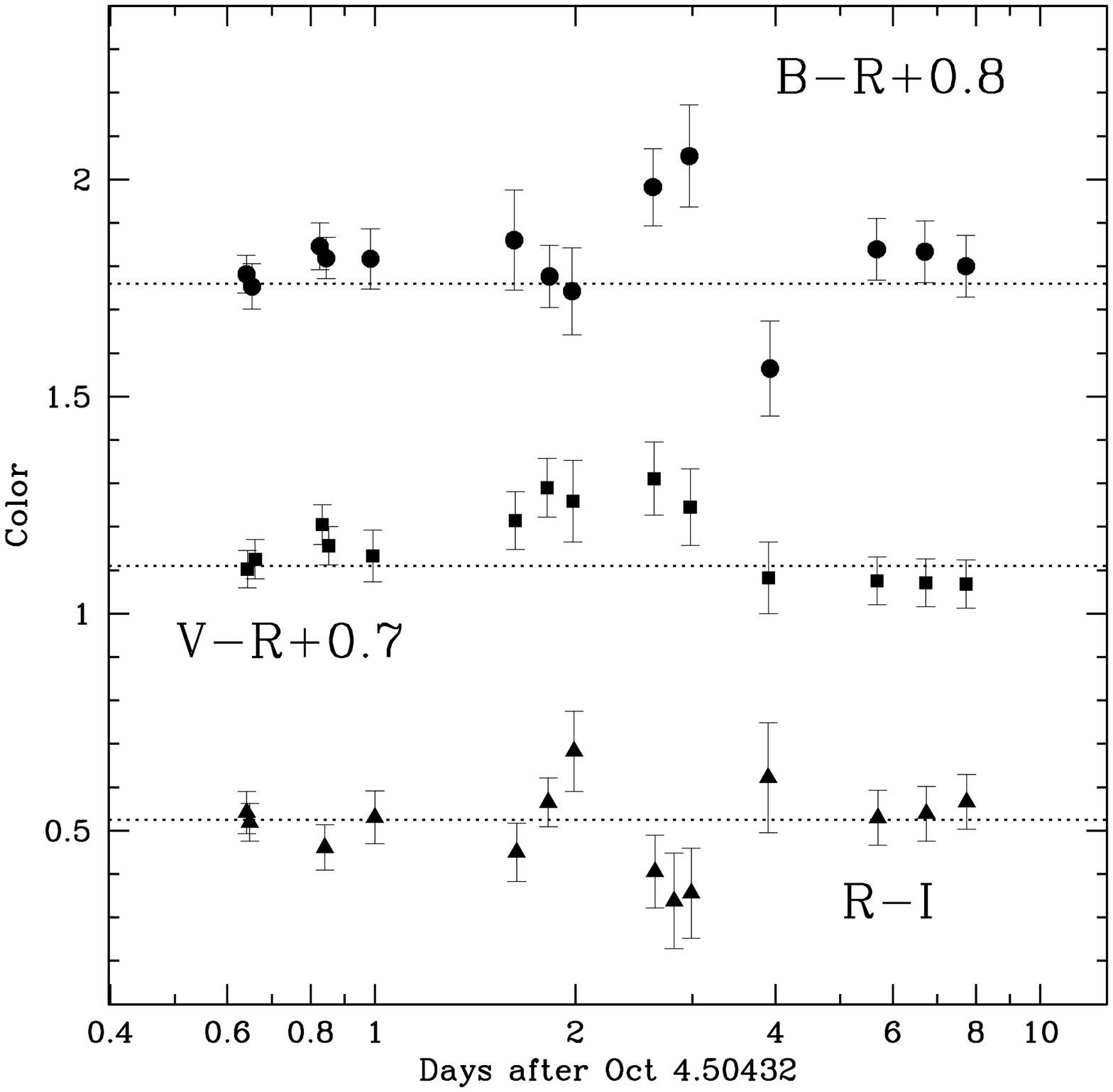}
\caption{Color evolution of the OT in $B-R$ (dots), $V-R$ (squares),
and $R-I$ (triangles). The dotted lines are the average color of the
first two points.
\label{fig_color}}
\end{figure}

The light curves for various colors of the OT are plotted in
Fig.~\ref{fig_color}, in which the color changes are more readily
seen. For instance, note the large change in $B-R$ between night 3 and
night 4. The OT became bluer, whereas $V-R$ changed only mildly and
$R-I$ seems to indicate a redder color. Evidently the shape of the
spectral energy distribution (SED) changed significantly $1.5-4$ days
after the burst.  The Magellan data taken $6-8\;$days after the burst
indicate that the afterglow has then returned to approximately the
same color it had during the first night. We await with interest the
results of the final analysis of the multi-band data taken by other
observers.

GRB 021004 is located at Galactic coordinates $l=114\arcdeg\!\!.9187,
b=-43\arcdeg\!\!.5615$. To remove the effects of the Galactic
interstellar extinction we used the reddening map of \citet{sfd98}
which yields $E(B-V)=0.06$. This corresponds to expected values of
Galactic extinction ranging from $A_I=0.12$ to $A_U=0.33$.

\begin{figure}[h]
\plotone{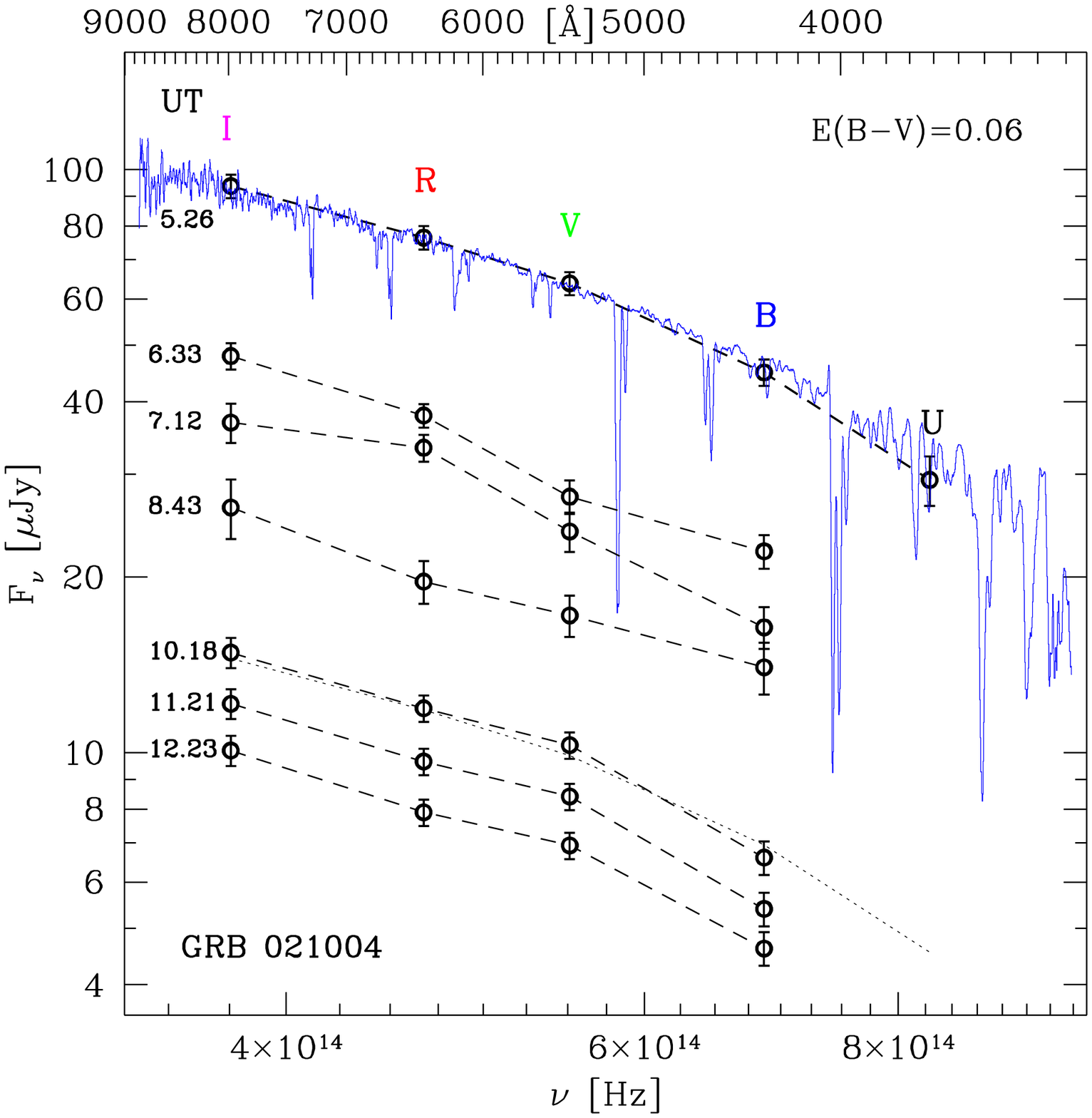}
\caption{Spectral energy distribution (SED) of the optical afterglow
of GRB 021004 at various times (indicated on the left side of each
SED).  We superimposed a spectrum obtained nearly simultaneously with
our photometry \citep{tom02}. The SED from UT 5.26 is shown as the
dotted line on top of SED from UT 10.18.
\label{fig_spec}}
\end{figure}

We synthesized the $UBVRI$ spectrum for the first night and $BVRI$
spectra for later nights from our data by interpolating the magnitudes
to a common time for the first night and using our best, most
closely-spaced measurements for the other nights
(Fig.~\ref{fig_spec}). We converted the magnitudes to fluxes using the
effective frequencies and normalizations of \citet{fsi95}.  These
conversions are accurate to about 4\%, so to account for the
calibration errors we added a 4\% error (7\% for the $U$-band) in
quadrature to the statistical error in each flux.

There are several important things to notice in Fig.~\ref{fig_spec}.
First, the SED on Oct. 5.26 is clearly curved at the blue end.  The
energy distribution of GRBs is usually a power law in the optical
domain (a straight line in Fig.~\ref{fig_spec}; see Garnavich et
al. 2002 for a striking example). This is clearly not the case here.
Our photometry is in very good agreement with the (independently
calibrated) high \emph{Signal/Noise} spectrum also displayed
\citep{tom02}.  We cannot exclude extra reddening in the host galaxy
of the GRB.
Another feature, coupled to the color variations discussed above, is
the evolution of the shape of the SED, with a most drastic change
between UT 7.12 and 8.43. After UT 10.0 the SED comes back to
approximately the same shape it had the first night.

The following picture emerges from all of our data: on top of the
``normal'' decay of the afterglow there seems to be a $30-40$\% bump,
fairly well localized in energy, propagating from the $I$-band $1.5-2$
days after the burst, through the $VR$-bands $2.5-3$ days after the
burst, to the $B$-band $4$ days after the burst. After $\sim 6$ days
the energy distribution comes back to the one it had on the first
night. This could be due, as suggested by Rhoads et al. (2002), to
``arrival of fresh energy at the blast wave external shock, carried by
slow ejecta'', but a detailed discussion is beyond the scope of this
paper.

\section{Short term variations}
\label{sec_day1}

\begin{figure}
\plotone{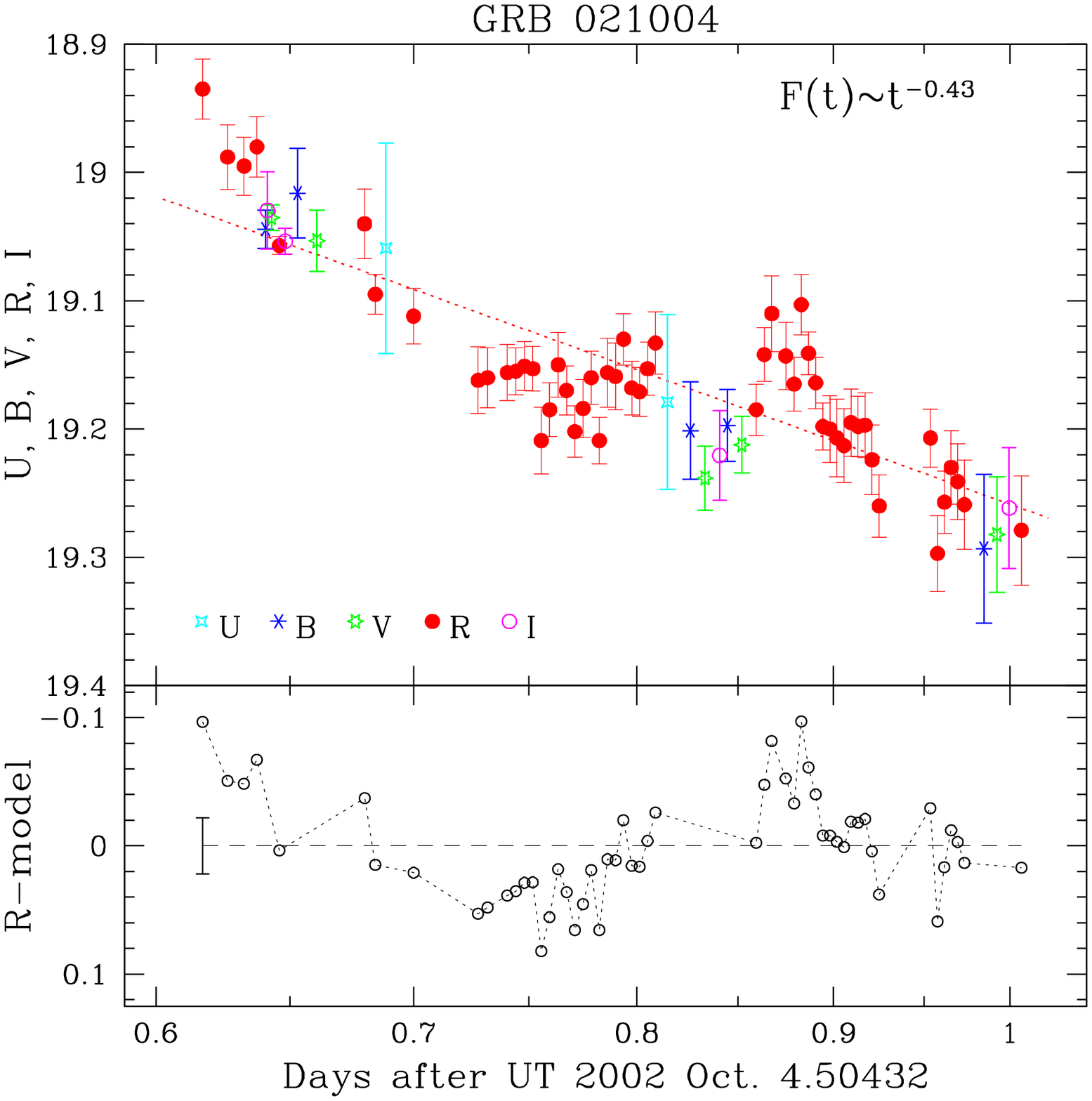}
\caption{{\it Upper panel} Light curve of the optical afterglow of GRB
021004 during the first night. A power law has been fitted to the data
(dotted line).  The $U$, $B$, $V$ and $I$ data have been shifted (see text
for details). {\it Lower panel} Residuals between this power law model
and the data.  The error bar on the left is typical for non-variable
stars with magnitude similar to the GRB ($rms \sim 0.022$ mag).
\label{fig_day1}}
\end{figure}

Encouraged by the detection of short-term variations observed in the
optical afterglow of GRB 011211 \citep{holland02} we decided to spend
most of an entire night monitoring this burst in order to search for
short term variations.  Our data, starting about 14.8 hours after the
burst, showed that the fading did indeed continue \citep{gcn1576}.
Then, at $\sim 18$ hours, fading stalled.  However, the OT was not
constant in brightness during this time. We observed short-term
variability on several time scales (see Fig.~\ref{fig_day1}).  This
has been confirmed independently by \citet{gcn1578}.  We fitted a
power law to our first night $UBVRI$ data; this yielded a decay slope
of $0.43$. A power law is obviously an inadequate description of the
OT but it allows us to interpolate the $UBVI$ magnitudes and transform
them into an ``equivalent'' $R$ magnitude\footnote{The power law index
is different from the index $\alpha_1$ derived in
Sect.~\ref{sec_time}. This is because in this case we fitted our first
night data only.}.

Inspection of Fig.~\ref{fig_day1} reveals that there is a trend over a
few hours (down then up then down again), upon which is superimposed a
short term, $\sim 10$\% variability with a time-scale of 15-30
minutes. This short-term variability is most obvious between 0.75 and
0.8 day, and again between 0.85 and 0.9 day. Such variations might
also be present around 0.95 day. We confirmed these variations with
three photometry packages.  Several nearby comparison stars with
brightness comparable to the OT were found to have rms of 0.02-0.025
mag, with no correlated variability present in their light curves. We
are therefore confident that the short-term variability is real and is
not an artifact of data reduction or statistical fluctuations.

It is only the second time that short-term variations have been
observed in a GRB optical afterglow (after GRB 011211: Holland et
al. 2002c). Our current data set is much better sampled and allows for
better study of this phenomenon. In at least one case, despite very
well sampled light curve, no variations larger than $\sim 0.02\;$mag
were present (GRB 990510: Stanek et al. 1999). Possible microlensing
has been seen in one instance \citep{gls00}, however it is probably
not the cause of the variations we are seeing here.

\section{Conclusion}
\label{sec_concl}

Several kinds of models can explain the early ($\sim 0.1$ day) and
later bumps seen on the light curve. For instance \citet{wl00} showed
that density fluctuations in the interstellar medium (ISM) surrounding
the GRB can induce significant photometric variability.  However other
mechanisms might be acting: several models have been proposed
specifically for this burst.

\citet{lazzati02} consider density fluctuations, either due to a
clumpy medium or a wind environment. They favor a clumpy ISM with a
density contrast of order 10. Their models reproduce fairly well the
first and second bumps in the light curve. They did not try to model
later bumps or short-term variability.  \citet{npg02} considered both
a variable density profile (clumpy ISM or stellar wind) and variable
energy in the blast wave (refreshed shocks or angular dependence of
jet).  Both types of models seem to reproduce the $R$-band light curve
fairly well although they do prefer the ``patchy shell'' model. A
possible shortcoming of these models is that the shape of the spectral
energy distribution is not supposed to change, whereas we do observe a
clear color change.  \citet{kz02} explain the first re-brightening (at
$\sim 0.1$ days) with a reverse shock. However, subsequent to this
first bump, their light curve is perfectly smooth they can not explain
the later bumps.  Other elaborations of this model would have to be
included (such as local energy variations or density inhomogeneities).

In conclusion, all models can account reasonably well for the first
bump on the light curve but no model yet provides a complete picture
of this optical afterglow.  Accurate modeling of the later bumps and
short term ``wiggles'' (see Fig.~\ref{fig_day1}) will require more
detailed work. Furthermore the changes in the energy distribution will
have to be taken into account by future models.


\acknowledgments

We thank the HETE team, Scott Barthelmy and the GRB Coordinates
Network (GCN) for the quick turnaround in providing precise GRB
positions to the astronomical community.  The amount of data and
excitement generated by this unusual burst owe much to the prompt
discovery of its optical afterglow and we thank D. Fox for his very
early efforts.  We thank all the observers who provided their data
through the GCN.  We thank D.~Bennett and K.~Cook for help with the
Boyden observations.  We also thank Jeremy Heyl, Rosalba Perna
and Stephen Holland for
useful discussions.  DB acknowledges support from NSF grant
AST-9979812.  JNW is supported by an NSF Astronomy \& Astrophysics
Postdoctoral Fellowship, under grant AST-0104347.  PMG
acknowledges support from the NASA LTSA grant NAG5-9364.  Support for
BJM 
was provided by NASA through a Hubble Fellowship grant from the Space
Telescope Science Institute, which is operated by the Association of
Universities for Research in Astronomy, Incorporated, under NASA
contract NAS5-26555.  DS acknowledges the support of a Smithsonian
Astrophysical Observatory Clay Fellowship.

\clearpage

\begin{deluxetable}{ r r r l c r l }
\tabletypesize{\scriptsize}
\tablecaption{Photometric data \label{tbl_data}}
\tablewidth{0pt}
\tablehead{
\colhead{UT date} & \colhead{$m$}  & \colhead{$\sigma_m$} & %
\colhead{image} & \colhead{filter} & \colhead{$t_{exp}$ (s)}  & %
\colhead{Telescope} 
}
\startdata
  5.1929 &   19.727 &  0.082 & ff1012   & U  &   600.0 & FLWO 48"       \\
  5.3191 &   19.847 &  0.068 & ff1037   & U  &   900.0 & FLWO 48"       \\
  5.1450 &   20.036 &  0.020 & grb\_b2  & B  &   100.0 & CTIO 4m        \\
  5.1574 &   20.008 &  0.035 & ff1007   & B  &   600.0 & FLWO 48"       \\
  5.3304 &   20.193 &  0.038 & ff1038   & B  &   600.0 & FLWO 48"       \\
  5.3489 &   20.189 &  0.028 & ff1041   & B  &   600.0 & FLWO 48"       \\
  5.4890 &   20.285 &  0.058 & ff1067   & B  &   600.0 & FLWO 48"       \\
  6.1226 &   20.899 &  0.109 & ff1103   & B  &   600.0 & FLWO 48"       \\
  6.3331 &   20.907 &  0.060 & ff1109   & B  &   900.0 & FLWO 48"       \\
  6.4817 &   21.021 &  0.093 & ff1111   & B  &   600.0 & FLWO 48"       \\
  7.1203 &   21.233 &  0.080 & ff1202   & B  &   900.0 & FLWO 48"       \\
  7.4706 &   21.388 &  0.111 & ff1212   & B  &   900.0 & FLWO 48"       \\
  8.4289 &   21.404 &  0.103 & ff1307   & B  &  1200.0 & FLWO 48"       \\
 10.1807 &   22.220 &  0.060 & gg1001   & B  &   600.0 & Magellan 6.5m  \\
 11.2102 &   22.440 &  0.060 & gg1101   & B  &   600.0 & Magellan 6.5m  \\
 12.2311 &   22.610 &  0.060 & gg1201   & B  &   600.0 & Magellan 6.5m  \\
  5.1473 &   19.458 &  0.020 & grb\_v2  & V  &   100.0 & CTIO 4m        \\
  5.1650 &   19.476 &  0.024 & ff1008   & V  &   600.0 & FLWO 48"       \\
  5.3377 &   19.661 &  0.025 & ff1039   & V  &   600.0 & FLWO 48"       \\
  5.3562 &   19.635 &  0.022 & ff1042   & V  &   600.0 & FLWO 48"       \\
  5.4967 &   19.705 &  0.045 & ff1068   & V  &   600.0 & FLWO 48"       \\
  6.1299 &   20.352 &  0.054 & ff1104   & V  &   600.0 & FLWO 48"       \\
  6.3185 &   20.509 &  0.056 & ff1107   & V  &   600.0 & FLWO 48"       \\
  6.4890 &   20.638 &  0.086 & ff1112   & V  &   600.0 & FLWO 48"       \\
  7.1311 &   20.661 &  0.075 & ff1203   & V  &   600.0 & FLWO 48"       \\
  7.4813 &   20.682 &  0.080 & ff1213   & V  &   900.0 & FLWO 48"       \\
  8.4081 &   21.014 &  0.073 & ff1306   & V  &  1200.0 & FLWO 48"       \\
 10.1882 &   21.550 &  0.040 & gg1002   & V  &   600.0 & Magellan 6.5m  \\
 11.2322 &   21.770 &  0.040 & gg1102   & V  &   600.0 & Magellan 6.5m  \\
 12.2385 &   21.980 &  0.040 & gg1202   & V  &   600.0 & Magellan 6.5m  \\
  4.9200 &   18.250 &  0.150 & sa       & R  &   138.0 & Boyden 1.52m   \\
  5.1213 &   18.935 &  0.027 & ff1001   & R  &   300.0 & FLWO 48"       \\
  5.1306 &   18.988 &  0.029 & ff1002   & R  &   300.0 & FLWO 48"       \\
  5.1368 &   18.995 &  0.026 & ff1003   & R  &   300.0 & FLWO 48"       \\
  5.1417 &   18.980 &  0.027 & ff1004   & R  &   300.0 & FLWO 48"       \\
  5.1503 &   19.057 &  0.020 & grb\_r2  & R  &   100.0 & CTIO 4m        \\
  5.1841 &   19.040 &  0.031 & ff1010   & R  &   300.0 & FLWO 48"       \\
  5.1885 &   19.095 &  0.020 & ff1011   & R  &   300.0 & FLWO 48"       \\
  5.2043 &   19.112 &  0.025 & ff1014   & R  &   300.0 & FLWO 48"       \\
  5.2319 &   19.162 &  0.030 & ff1015   & R  &   300.0 & FLWO 48"       \\
  5.2360 &   19.160 &  0.027 & ff1016   & R  &   300.0 & FLWO 48"       \\
  5.2447 &   19.156 &  0.025 & ff1018   & R  &   300.0 & FLWO 48"       \\
  5.2485 &   19.155 &  0.021 & ff1019   & R  &   300.0 & FLWO 48"       \\
  5.2523 &   19.151 &  0.022 & ff1020   & R  &   300.0 & FLWO 48"       \\
  5.2561 &   19.153 &  0.020 & ff1021   & R  &   300.0 & FLWO 48"       \\
  5.2599 &   19.209 &  0.030 & ff1022   & R  &   300.0 & FLWO 48"       \\
  5.2637 &   19.185 &  0.024 & ff1023   & R  &   300.0 & FLWO 48"       \\
  5.2676 &   19.150 &  0.029 & ff1024   & R  &   300.0 & FLWO 48"       \\
  5.2714 &   19.170 &  0.022 & ff1025   & R  &   300.0 & FLWO 48"       \\
  5.2752 &   19.202 &  0.023 & ff1026   & R  &   300.0 & FLWO 48"       \\
  5.2790 &   19.184 &  0.026 & ff1027   & R  &   300.0 & FLWO 48"       \\
  5.2828 &   19.160 &  0.024 & ff1028   & R  &   300.0 & FLWO 48"       \\
  5.2866 &   19.209 &  0.021 & ff1029   & R  &   300.0 & FLWO 48"       \\
  5.2904 &   19.156 &  0.031 & ff1030   & R  &   300.0 & FLWO 48"       \\
  5.2942 &   19.159 &  0.030 & ff1031   & R  &   300.0 & FLWO 48"       \\
  5.2980 &   19.130 &  0.023 & ff1032   & R  &   300.0 & FLWO 48"       \\
  5.3019 &   19.168 &  0.024 & ff1033   & R  &   300.0 & FLWO 48"       \\
  5.3057 &   19.171 &  0.022 & ff1034   & R  &   300.0 & FLWO 48"       \\
  5.3095 &   19.153 &  0.024 & ff1035   & R  &   300.0 & FLWO 48"       \\
  5.3133 &   19.133 &  0.028 & ff1036   & R  &   300.0 & FLWO 48"       \\
  5.3636 &   19.185 &  0.023 & ff1043   & R  &   300.0 & FLWO 48"       \\
  5.3677 &   19.142 &  0.024 & ff1044   & R  &   300.0 & FLWO 48"       \\
  5.3716 &   19.110 &  0.034 & ff1045   & R  &   300.0 & FLWO 48"       \\
  5.3789 &   19.143 &  0.030 & ff1046   & R  &   300.0 & FLWO 48"       \\
  5.3833 &   19.165 &  0.024 & ff1047   & R  &   300.0 & FLWO 48"       \\
  5.3871 &   19.103 &  0.027 & ff1048   & R  &   300.0 & FLWO 48"       \\
  5.3909 &   19.141 &  0.019 & ff1049   & R  &   300.0 & FLWO 48"       \\
  5.3947 &   19.164 &  0.023 & ff1050   & R  &   300.0 & FLWO 48"       \\
  5.3985 &   19.198 &  0.021 & ff1051   & R  &   300.0 & FLWO 48"       \\
  5.4023 &   19.200 &  0.030 & ff1052   & R  &   300.0 & FLWO 48"       \\
  5.4061 &   19.207 &  0.035 & ff1053   & R  &   300.0 & FLWO 48"       \\
  5.4099 &   19.213 &  0.033 & ff1054   & R  &   300.0 & FLWO 48"       \\
  5.4137 &   19.195 &  0.030 & ff1055   & R  &   300.0 & FLWO 48"       \\
  5.4176 &   19.198 &  0.027 & ff1056   & R  &   300.0 & FLWO 48"       \\
  5.4214 &   19.197 &  0.029 & ff1057   & R  &   300.0 & FLWO 48"       \\
  5.4252 &   19.224 &  0.031 & ff1058   & R  &   300.0 & FLWO 48"       \\
  5.4292 &   19.260 &  0.028 & ff1059   & R  &   300.0 & FLWO 48"       \\
  5.4580 &   19.207 &  0.026 & ff1060   & R  &   300.0 & FLWO 48"       \\
  5.4620 &   19.297 &  0.034 & ff1061   & R  &   300.0 & FLWO 48"       \\
  5.4660 &   19.257 &  0.028 & ff1062   & R  &   300.0 & FLWO 48"       \\
  5.4699 &   19.230 &  0.033 & ff1063   & R  &   300.0 & FLWO 48"       \\
  5.4737 &   19.241 &  0.034 & ff1064   & R  &   300.0 & FLWO 48"       \\
  5.4775 &   19.259 &  0.040 & ff1065   & R  &   300.0 & FLWO 48"       \\
  5.5113 &   19.279 &  0.049 & ff1070   & R  &   300.0 & FLWO 48"       \\
  6.1075 &   19.868 &  0.053 & ff1101   & R  &   600.0 & FLWO 48"       \\
  6.1150 &   19.841 &  0.048 & ff1102   & R  &   600.0 & FLWO 48"       \\
  6.3111 &   19.913 &  0.032 & ff1106   & R  &   600.0 & FLWO 48"       \\
  6.4744 &   20.078 &  0.052 & ff1110   & R  &   600.0 & FLWO 48"       \\
  7.1090 &   20.051 &  0.041 & ff1201   & R  &   900.0 & FLWO 48"       \\
  7.3084 &   20.051 &  0.111 & ff1205   & R  &   900.0 & FLWO 48"       \\
  7.3671 &   20.153 &  0.054 & ff1207   & R  &   600.0 & FLWO 48"       \\
  7.3806 &   20.129 &  0.051 & ff1208   & R  &   600.0 & FLWO 48"       \\
  7.3879 &   20.167 &  0.050 & ff1209   & R  &   600.0 & FLWO 48"       \\
  7.3954 &   20.124 &  0.034 & ff1210   & R  &   600.0 & FLWO 48"       \\
  7.4633 &   20.132 &  0.063 & ff1211   & R  &   600.0 & FLWO 48"       \\
  7.5029 &   20.144 &  0.100 & ff1215   & R  &   300.0 & FLWO 48"       \\
  8.3068 &   20.436 &  0.131 & ff1302   & R  &   900.0 & FLWO 48"       \\
  8.3918 &   20.625 &  0.082 & ff1304   & R  &   600.0 & FLWO 48"       \\
  9.3453 &   20.858 &  0.044 & ff1401   & R  &  1200.0 & FLWO 48"       \\
  9.3596 &   20.883 &  0.050 & ff1402   & R  &  1200.0 & FLWO 48"       \\
 10.1960 &   21.168 &  0.040 & gg1003   & R  &   600.0 & Magellan 6.5m  \\
 10.2112 &   21.153 &  0.040 & gg1005   & R  &   300.0 & Magellan 6.5m  \\
 10.2150 &   21.180 &  0.040 & gg1006   & R  &   300.0 & Magellan 6.5m  \\
 10.2187 &   21.170 &  0.040 & gg1007   & R  &   300.0 & Magellan 6.5m  \\
 10.2225 &   21.180 &  0.040 & gg1008   & R  &   300.0 & Magellan 6.5m  \\
 11.2398 &   21.396 &  0.040 & gg1103   & R  &   600.0 & Magellan 6.5m  \\
 11.2560 &   21.380 &  0.040 & gg1105   & R  &   600.0 & Magellan 6.5m  \\
 12.2458 &   21.614 &  0.040 & gg1203   & R  &   600.0 & Magellan 6.5m  \\
 13.1176 &   21.842 &  0.040 & gg1301   & R  &   600.0 & Magellan 6.5m  \\
 13.1249 &   21.844 &  0.040 & gg1302   & R  &   600.0 & Magellan 6.5m  \\
 25.5000 &   23.390 &  0.120 & vatt     & R  &   600.0 & VATT 1.8m      \\
  5.1457 &   18.513 &  0.030 & ff1005   & I  &   300.0 & FLWO 48"       \\
  5.1526 &   18.537 &  0.020 & grb\_i2  & I  &   100.0 & CTIO 4m        \\
  5.3450 &   18.704 &  0.035 & ff1040   & I  &   300.0 & FLWO 48"       \\
  5.5040 &   18.745 &  0.047 & ff1069   & I  &   600.0 & FLWO 48"       \\
  6.1372 &   19.388 &  0.055 & ff1105   & I  &   600.0 & FLWO 48"       \\
  6.3258 &   19.359 &  0.041 & ff1108   & I  &   600.0 & FLWO 48"       \\
  6.4963 &   19.396 &  0.084 & ff1113   & I  &   600.0 & FLWO 48"       \\
  7.1384 &   19.644 &  0.074 & ff1204   & I  &   600.0 & FLWO 48"       \\
  7.3192 &   19.715 &  0.103 & ff1206   & I  &   300.0 & FLWO 48"       \\
  7.4921 &   19.784 &  0.096 & ff1214   & I  &   900.0 & FLWO 48"       \\
  8.4042 &   20.008 &  0.121 & ff1305   & I  &   600.0 & FLWO 48"       \\
 10.2034 &   20.630 &  0.050 & gg1004   & I  &   600.0 & Magellan 6.5m  \\
 11.2470 &   20.850 &  0.050 & gg1104   & I  &   600.0 & Magellan 6.5m  \\
 12.2531 &   21.050 &  0.050 & gg1204   & I  &   600.0 & Magellan 6.5m  \\
\enddata
\end{deluxetable}

\end{document}